\documentclass[aps,prl,10pt,twocolumn,showpacs,superscriptaddress]{revtex4-1}

\usepackage{graphicx}
\usepackage{dcolumn}
\usepackage{bm}
\usepackage{times}
\usepackage{amsmath}
\usepackage{amssymb}
\usepackage{subfigure}
\usepackage{epsfig}

\begin{document}

\title{Charge-order-induced ferroelectricity in LaVO$_{3}$/SrVO$_{3}$ superlattices}
\author{Se Young Park}
\email{sypark@physics.rutgers.edu}
\affiliation{Department of Physics \& Astronomy, Rutgers University, Piscataway, NJ 08854, USA}
\author{Anil Kumar}
\affiliation{Theoretical Division, Los Alamos National Laboratory, NM 87545, USA}
\author{Karin M. Rabe}
\affiliation{Department of Physics \& Astronomy, Rutgers University, Piscataway, NJ 08854, USA}

\begin{abstract}
The structure and properties of the 1:1 superlattice of LaVO$_{3}$ and SrVO$_{3}$ are investigated with a first-principles density-functional-theory-plus-$U$ (DFT+$U$) method. The lowest energy states are antiferromagnetic charge-ordered Mott-insulating phases. In one of these insulating phases, layered charge ordering combines with the layered cation ordering to produce a polar structure with nonzero spontaneous polarization normal to the interfaces. This polarization is produced by electron transfer between the V$^{3+}$ and V$^{4+}$ layers, and is comparable to that of conventional ferroelectrics. The energy of this polar state relative to the nonpolar ground state is only 3 meV per vanadium. Under tensile strain, this energy difference can be further reduced, suggesting that the polar phase can be induced by applied electric field, yielding an antiferroelectric double-hysteresis loop. If the system does not switch back to the nonpolar state on removal of the field, a ferroelectric-type hysteresis loop could be observed. 
\end{abstract}

%

\maketitle


The investigation of novel mechanisms for ferroelectricity has attracted much interest in recent years, especially in the search for ferroelectrics with properties, such as ferromagnetism, that are contraindicated by the mechanism driving ferroelectricity in the prototypical perovskite titanates \cite{Hill00}. 
One approach, proposed by Khomskii \cite{Efremov04,vandenBrink08} is to combine two symmetry-breaking orderings, neither of which separately lift inversion symmetry, to generate a switchable polar structure \cite{Young16}.
The specific orderings discussed by Khomskii are site-centered charge ordering and bond-centered charge ordering. 
Ferroelectricity and multiferroelectricity induced by charge order have been proposed and reported in various magnetites, manganates, and charge transfer organic salts \cite{Efremov04,vandenBrink08,Kunihiko14}.
In some of these materials, the polarization is dominated by the electron transfer producing the charge order rather than by ionic displacements, leading to the term ``electronic ferroelectricity.'' \cite{Ishihara10,Kobayashi12,He16}

In perovskite transition metal (TM) oxide (ABO$_{3}$)$_{n}$(A$^{\prime}$BO$_{3}$)$_{m}$ (001) superlattices, the layered cation ordering lowers the symmetry from cubic to tetragonal and breaks up-down symmetry across BO$_{2}$ layers. Control of the TM $d$-orbital occupancy through choice of A cations and layer thicknesses to obtain a mixed valence leads to charge disproportionation and long-range charge ordering \cite{Imada98,Pentcheva07,Bristowe15,Krick16}. As we will discuss further below, layered charge ordering breaks the up-down symmetry across the AO and A$^{\prime}$O layers. Thus, the combination of these two symmetry-breaking orderings (TM site-centered charge ordering and layered cation ordering) can generate a switchable polar structure with polarization normal to the layers. 

A 1:1 superlattice composed of LaVO$_{3}$ and SrVO$_{3}$ is a promising candidate for this type of charge-order-induced ferroelectricity. The low temperature phases of orthorhombic LaVO$_{3}$ and cubic SrVO$_{3}$ are antiferromagnetic Mott-insulating and correlated metallic, respectively \cite{Rey90,Bordet93,Imada98,Yoshida05}. When they form the 1:1 superlattice, the average valence of the vanadium has a fractional value of +3.5. Due to the strong on-site Coulomb interaction, the Mott-insulating state with charge, orbital, and magnetic order would be preferred where the vanadiums disproportionate so that half of the vanadium cations have nominal valence V$^{3+}$ and the other half V$^{4+}$ \cite{Quan12,Pickett14}. 

In this paper, we use first-principles total energy calculations to investigate the low-energy charge-ordered phases of the (LaVO$_{3}$)$_{1}$(SrVO$_{3}$)$_{1}$ superlattice. We show that there are three competing low-energy phases with distinct charge-order patterns, one of which is a layered charge ordering, and analyze the density of states and local structure. The epitaxial strain dependence of the relative energies is computed. For the polar structure generated by the layered charge order, we compute and discuss the spontaneous polarization normal to the layers, and predict the stabilization of this phase by an applied electric field.

We performed first-principles density-functional-theory calculations with the generalized gradient approximation plus $U$ (GGA+$U$) method using the Vienna {\it ab-initio} simulation package \cite{Kresse96,Kresse99}. The Perdew-Becke-Erzenhof parametrization \cite{Perdew96} for the exchange-correlation functional and the rotationally invariant form of the on-site Coulomb interaction \cite{Liechtenstein95} are used with $U=4$ eV and $J=0.6$ eV for the vanadium $d$ states and $U_{f}=11$ eV and $J_{f}=0.68$ eV for the La $f$ states to shift the La $f$ bands away from the Fermi energy \cite{Czyzyk94}. We used the projector augmented wave method \cite{Blochl94} with an energy cutoff of 600 eV and $k$-point sampling on a $4\times 4\times 4$ grid with a $2\times 2\times 2$ unit cell chosen to accommodate the relevant octahedral rotation distortions and charge-order patterns. The ferroelectric polarization was calculated using the Berry phase method \cite{King-Smith93} with $6\times 6\times 4$ $k$-point grid.

\begin{figure}[htbp]
\begin{center}
\includegraphics[width=1\columnwidth, angle=-0]{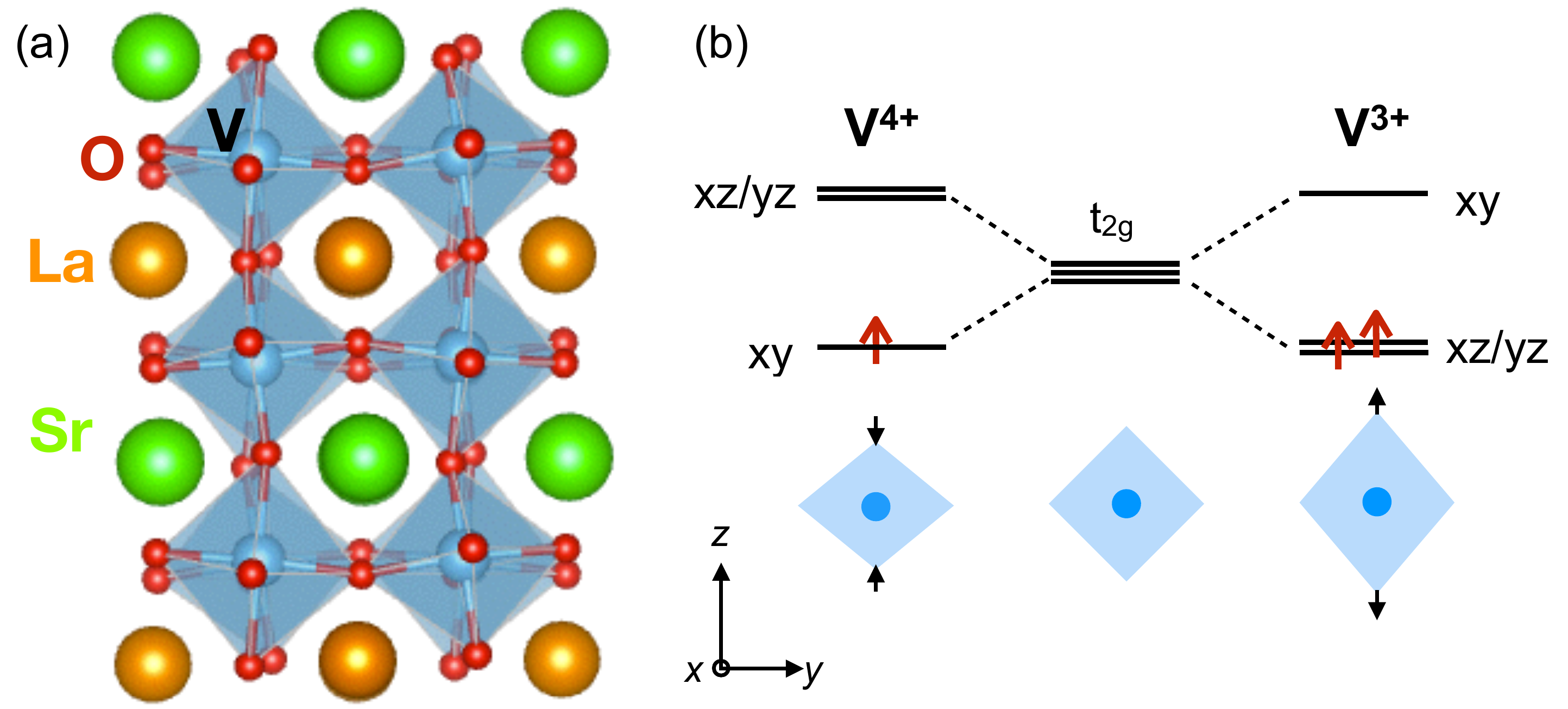}
\caption{(Color online) (a) Atomic arrangement in the (LaVO$_{3}$)$_{1}$/(SrVO$_{3}$)$_{1}$ superlattice. (b) The Jahn-Teller distortions and $t_{2g}$ level splittings for V$^{4+}$ (left) and V$^{3+}$ (right).}
\label{fig:str-jt}
\end{center}
\end{figure}

\begin{figure}[htbp]
\begin{center}
\includegraphics[width=0.9\columnwidth, angle=-0]{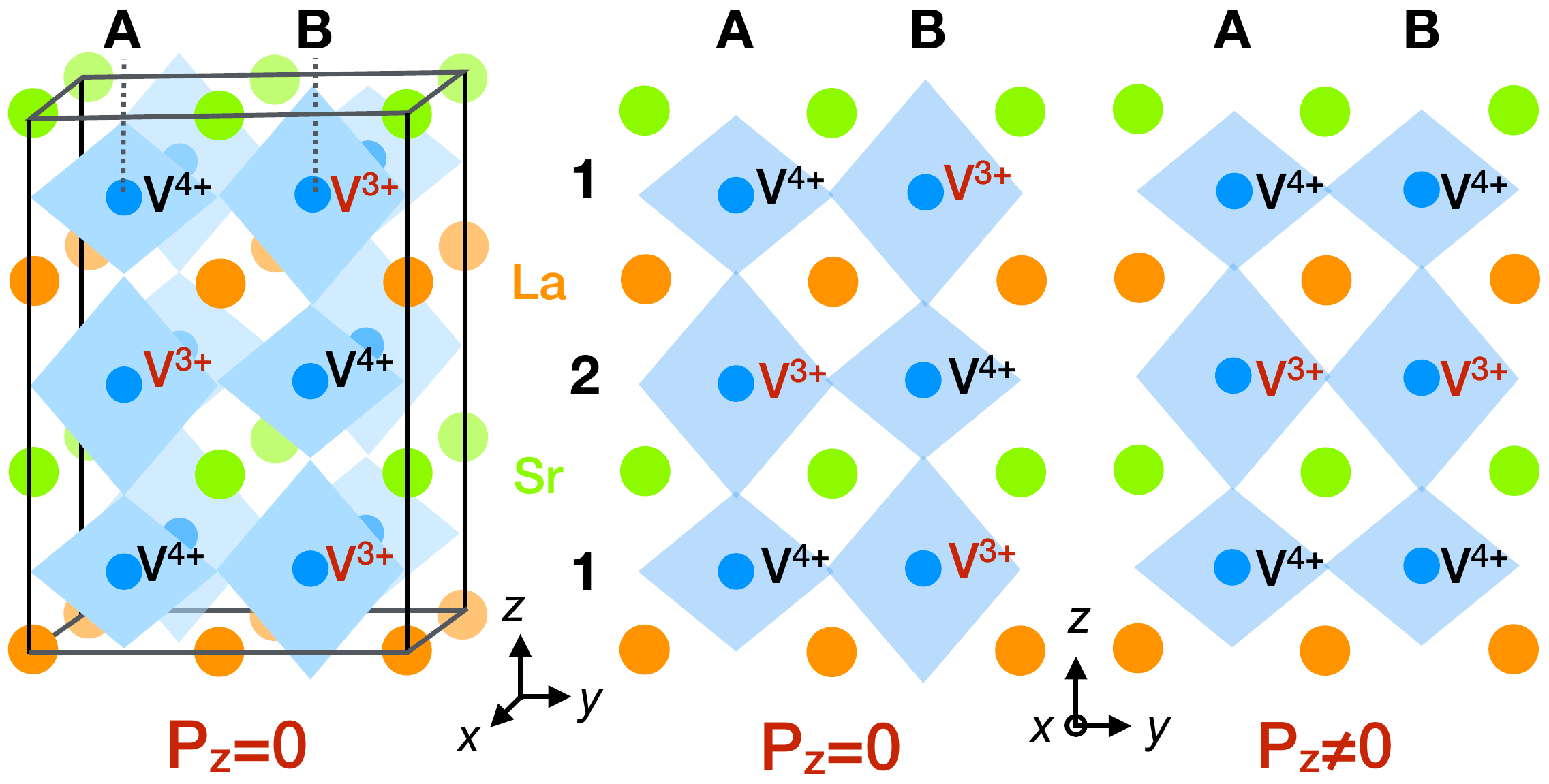}
\caption{(Color online) Schematics of the charge-order patterns for (LaVO$_{3}$)$_{1}$/(SrVO$_{3}$)$_{1}$ superlattice. For descriptive purposes the octahedral rotations and tilts are omitted and the Jahn-Teller distortions are exggerated.  The left, middle, right panels denote the checkerboard-type charge order with the $(\pi,\pi,\pi)$ ordering vector, the stripe charge order with the $(0,\pi,\pi)$ ordering vector, and the layered charge order with the $(0,0,\pi)$ ordering vector, respectively.}
\label{fig:copatterns}
\end{center}
\end{figure}

To determine the lowest-energy crystal structures and magnetic orderings of the 1:1 superlattice shown in Fig.~\ref{fig:str-jt}(a), we first observe that with average valence +3.5, the vanadium ions are expected to disproportionate to V$^{3+}$ and V$^{4+}$. Each valence state drives a corresponding Jahn-Teller distortion of its coordinating oxygen octahedron \cite{Varignon15,Varignon16}. As shown in Fig.~\ref{fig:str-jt} (b), the $d^{1}$ occupation for V$^{4+}$ favors a uniaxial contraction of the octahedron, while the $d^{2}$ occupation for V$^{3+}$ favors a uniaxial elongation \cite{Imada98}, with each site having fully spin-polarized electrons.

Given the 1:1 ratio of V$^{3+}$ to V$^{4+}$, we considered three cell-doubling charge-order patterns compatible with Jahn-Teller distortions along the $z$ axis, illustrated in Fig.~\ref{fig:copatterns}. In checkerboard charge ordering (CCO), which is the most common ordering in transition metal perovskite oxides \cite{Staub02}, all the nearest-neighbor B sites of V$^{4+}$ are V$^{3+}$ and vice versa. In stripe charge ordering (SCO), the nearest-neighbor B sites of V$^{4+}$ (V$^{3+}$) are V$^{4+}$ (V$^{3+}$) in the $x$ direction and V$^{3+}$ (V$^{4+}$) in the $y$ and $z$ directions, forming a 2D $y$-$z$ checkerboard with vanadium ions of a given valence forming lines along the $x$-direction. In layered charge ordering (LCO), planes of V$^{4+}$ and V$^{3+}$ alternate along the $z$-direction. 
Finally, we followed the stacking method \cite{Zhou14} to identify the oxygen octahedron rotation patterns of starting structures. Specifically, since bulk SrVO$_{3}$ has the cubic perovskite structure and bulk LaVO$_{3}$ has an orthorhombic $Pbnm$ structure with a $a^{-}a^{-}c^{+}$ rotation pattern \cite{Bordet93,Rey90}, we imposed the latter rotation pattern in two inequivalent orientations: one with the $c$ axis along the $z$-direction ($c$ oriented) and the other with the $c$ axis along the $x$-direction ($ab$ oriented). 

\begin{figure}[htbp]
\begin{center}
\includegraphics[width=0.9\columnwidth, angle=-0]{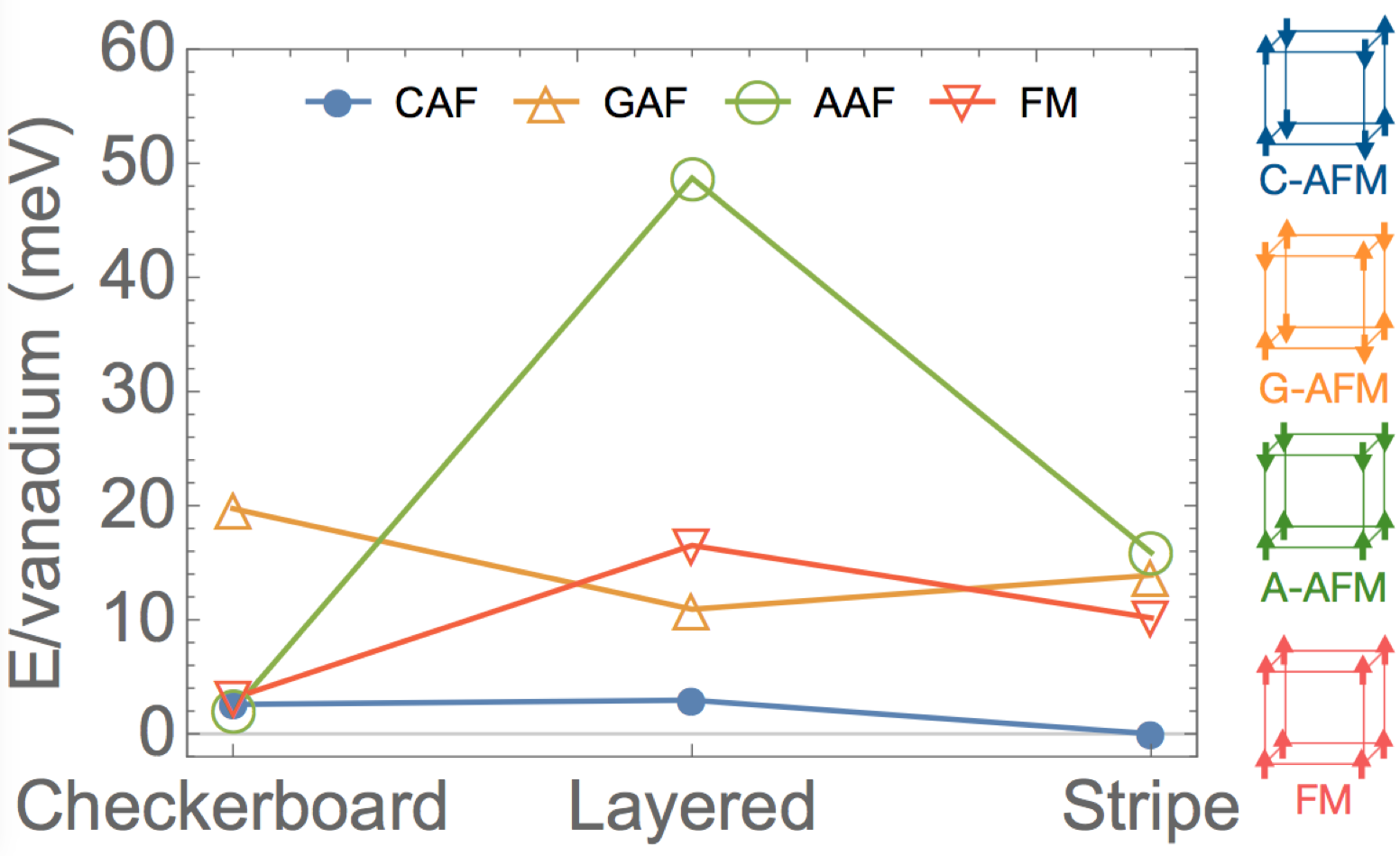}
\caption{(Color online) Total energy versus the charge ordering and magnetic ordering patterns. The symbols G, C, A, and FM represent the type of magnetic ordering of V-cations illustrated in the right column. }
\label{fig:envscomag}
\end{center}
\end{figure}

For each starting structure, we fully relaxed the lattice constants and atomic positions for each magnetic ordering. All relaxed structures are found to be insulating with a band gap larger than 1 eV. In all cases, the relaxed lattice has orthorhombic symmetry with lattice vectors close to cubic; for example, the lattice constants for LCO with C-AFM ordering are $a=5.65 \;\AA$, $b=5.62 \;\AA$, and $c=7.83 \; \AA$ for the 20 atom cell;  full structural and energetic information is given in the Supplementary Material. The energy difference between the two oxygen octahedron rotation patterns for a given charge and magnetic ordering is found to be small compared to the charge and magnetic ordering dependence; in what follows, we discuss $c$-oriented patterns.
Fig.~\ref{fig:envscomag} shows the total energies for the various magnetic orderings of each charge ordering pattern. The C-AFM ordering has the lowest energy for all three patterns, though for the CCO phase, the total energies of A-AFM and FM ordering are very close to that of C-AFM. It is remarkable that the energy differences between the lowest energy CCO, LCO, and SCO phases are quite small, with SCO computed to be lower in energy than LCO by only 3 meV per vanadium.

\begin{figure}[htbp]
\begin{center}
\includegraphics[width=0.9\columnwidth, angle=-0]{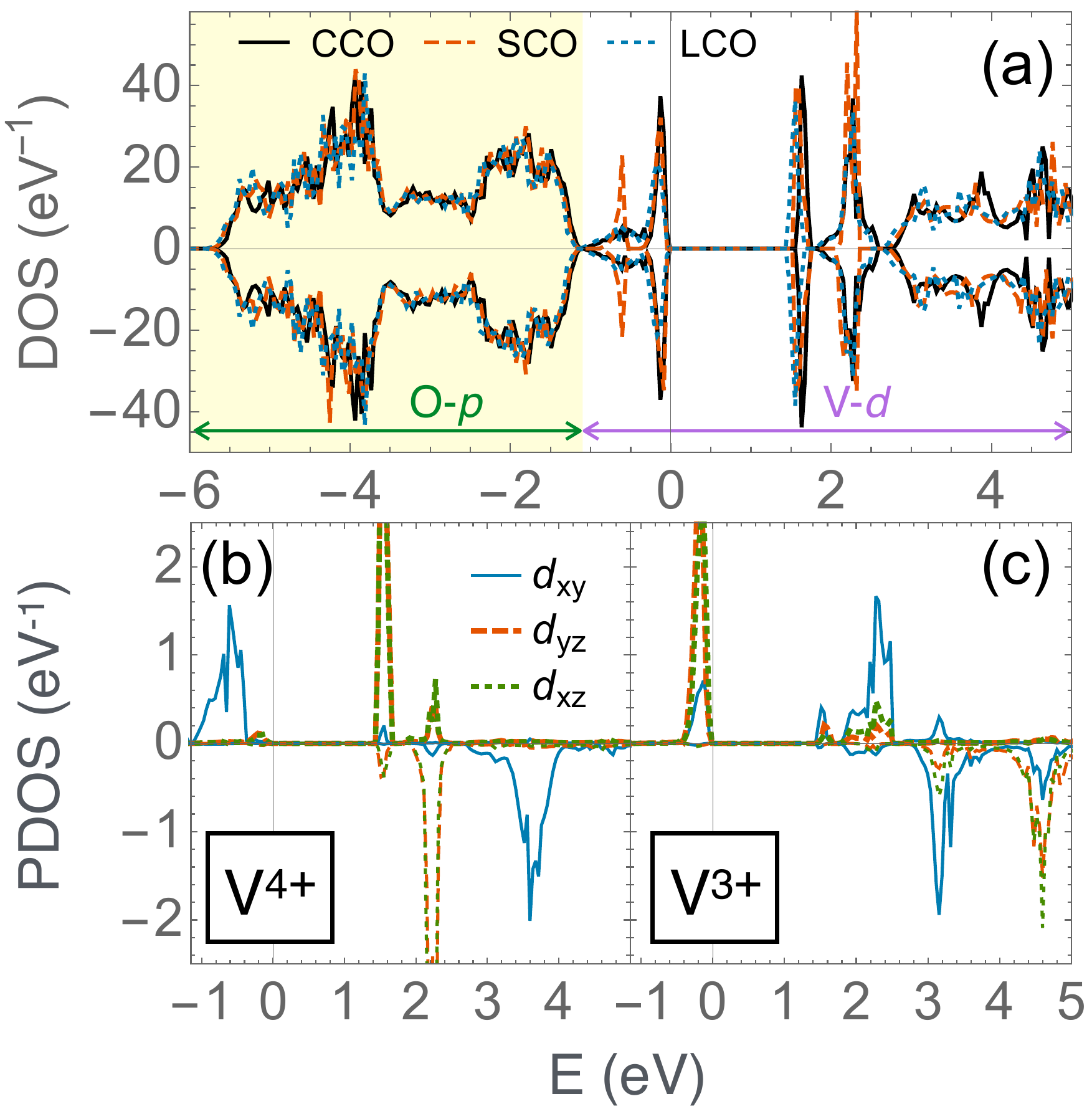}
\caption{(Color online) (a) Density of states of CCO (black solid line), SCO (red dashed line), and LCO (blue dotted line) phases with C-AFM ordering. The positive and negative sign of the vertical axis correspond to spin up and down states. The shaded (from -6 eV to -1.15 eV) and unshaded regions (from -1.15 eV to 5 eV) are oxygen $p$- and vanadium $d$-derived states, respectively. (b-c) PDOS of V-derived $t_{2g}$ states for the LCO phase with C-AFM ordering. (b) PDOS of the contracted octahedron (V$^{4+}$) with spin up polarization. (c) PDOS of the elongated octahedron (V$^{3+}$) with spin up polarization.}
\label{fig:dos-pdos}
\end{center}
\end{figure}

\begin{figure}[htbp]
\begin{center}
\includegraphics[width=0.9\columnwidth, angle=-0]{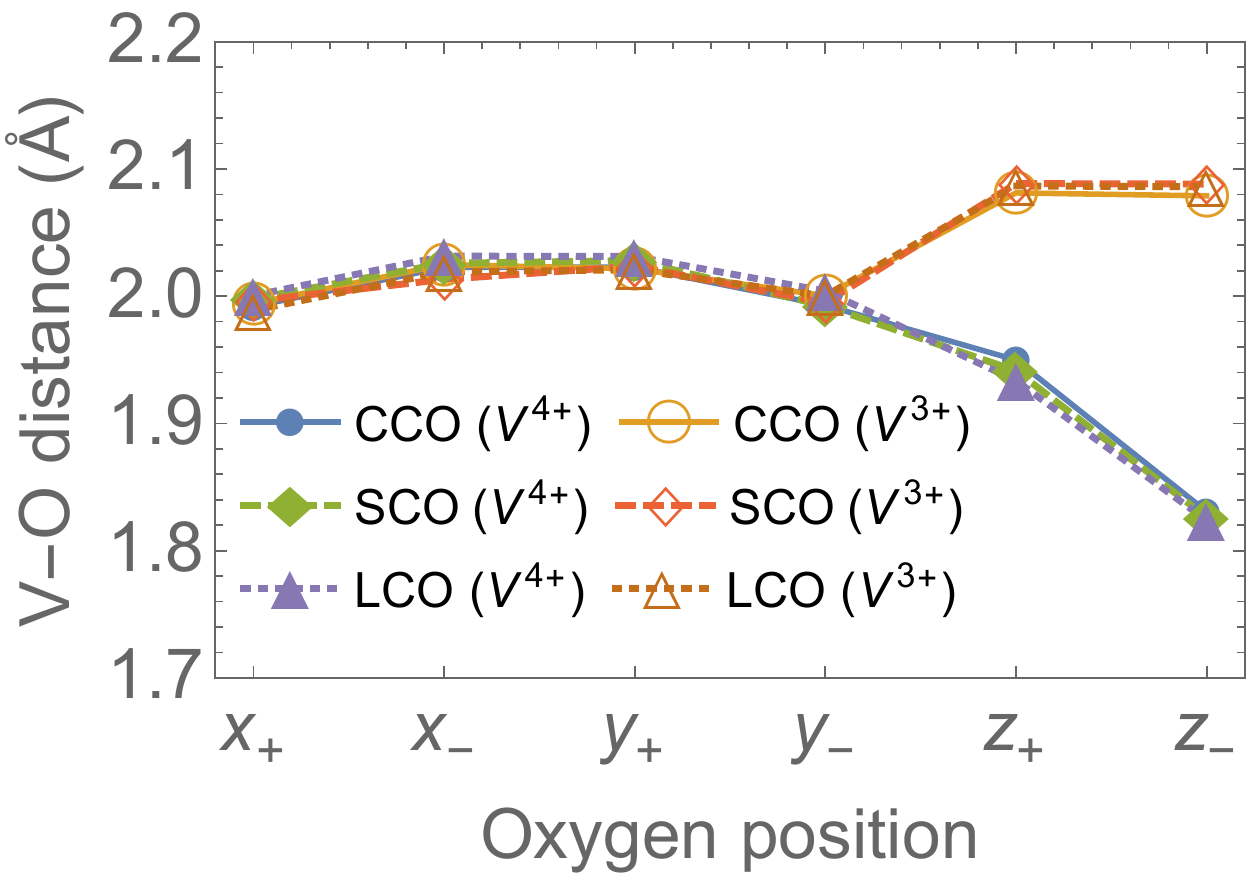}
\caption{(Color online) Distance between V and neighboring O atoms of $V^{4+}$ and $V^{3+}$ cations compared for CCO, SCO, and LCO phases. The $q_{+}$ and $q_{-}$ ($q=x,y,z$) denote the oxygen positions along the $q$-axis in the positive and negative direction, respectively.}
\label{fig:octvoldist}
\end{center}
\end{figure}

The densities of states (DOS) of the superlattice in CCO, SCO, and LCO phases with C-AFM ordering are shown in Fig.~\ref{fig:dos-pdos} (a). Consistent with the small energy difference between the phases, they are almost identical, the largest difference being in the occupied $d_{xy}$ states, with a sharp peak for SCO and a slightly broader distribution for CCO and LCO. The effect of the Jahn-Teller distortion can be seen from the projected density of states (PDOS), shown here for the LCO pattern (again, the corresponding PDOS for the other two phases are almost identical except for the occupied $d_{xy}$ states as noted above).  For the contracted octahedron (panel (b)) the occupied V-derived $t_{2g}$ states have $xy$ orbital character with full spin-polarization with $xz/yz$ states lying about 1.5 eV above the Fermi energy.  For the elongated octahedron (panel (c)), the $xz$ and $yz$ orbitals are occupied with empty $xy$ states. The site-projected magnetic moment for the contracted and elongated octahedra are 1.02$\mu_{B}$ and 1.78$\mu_{B}$, respectively. Since the occupied $d$-states are fully spin polarized, the site-projected magnetic moments are close to the $d$-occupancy, which supports the nominal V$^{4+}$/V$^{3+}$ valence states. 

\begin{figure}[htbp]
\begin{center}
\includegraphics[width=0.9\columnwidth, angle=-0]{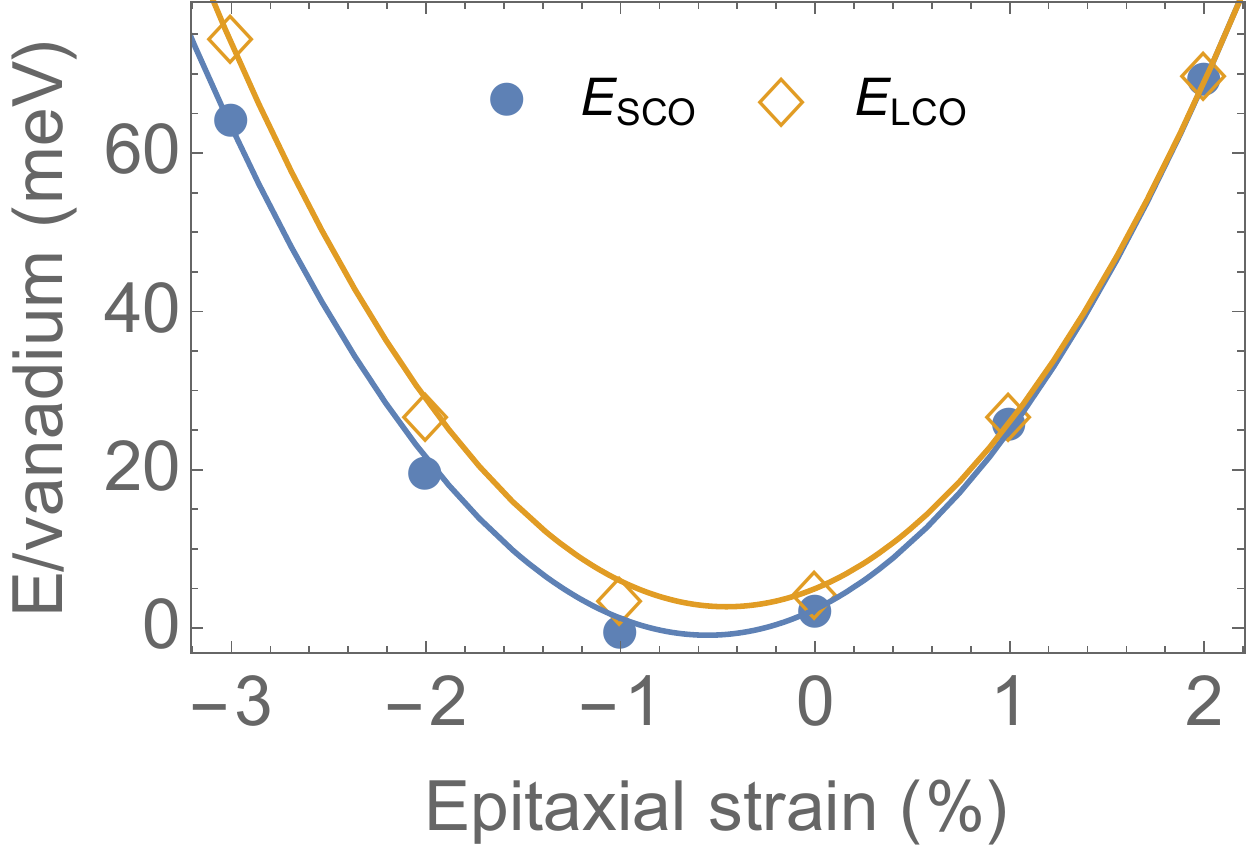}
\caption{(Color online) Total energy per vanadium as a function of epitaxial strain. The total energy is obtained with tetragonal unit cell where all the atomic position except in-plane lattice constants are fully relaxed. The zero strain is defined as the average lattice constant of the bulk LaVO$_{3}$ obtained from the GGA+$U$ calculation. }
\label{fig:envsstrain}
\end{center}
\end{figure}

In Fig.~\ref{fig:octvoldist} we compare the V-O bond lengths of the V$^{3+}$ and V$^{4+}$ octahedra of the three charge ordering patterns with C-AFM magnetic ordering. The in-plane bond lengths are close to equal in each case, with elongation and contraction in the $z$-direction for V$^{3+}$ and V$^{4+}$ octahedra, respectively, corresponding to the expected Jahn-Teller distortions. For each valence state, the shape of the octahedron is almost independent of the charge ordering pattern, suggesting that the distortion is a strongly local effect.

Next, we consider the symmetry and electric polarization of the various phases.
The layered cation ordering in the 1:1 superlattice lowers the cubic symmetry of the ideal perovskite structure to tetragonal $P4/mmm$ (\#123).
A $c$-oriented oxygen octahedron rotation pattern, combined with the layered cation ordering, produces a $Pmb2_{1}$ (\#26) structure with a small in-plane electric polarization resulting from non-cancellation of alternating A and A$^{\prime}$ in-plane displacements \cite{Benedek12,Bristowe15}; the computed value for the systems considered here is close to 4 $\mu$C/cm$^{2}$ (see supplementary material). 
A CCO charge ordering pattern lowers the symmetry further. The resulting $P2_{1}$ (\#4) space group contains a reflection in $z$ ($\bar{x}, y+\frac{1}{2},\bar{z}$) and thus the out-of-plane polarization is zero by symmetry. The SCO charge ordering in the $Pmb2_{1}$ (\#26) structure results in the space group $Pa$ (\#7) which contains  a reflection in $z$ ($x+1/2, y,\bar{z}$) and thus has zero out-of-plane polarization. 
In contrast, LCO charge ordering results in the space group $Pb$ (\#7) with ($\bar{x}, y+\frac{1}{2},z$), which allows a nonzero polarization along $z$. 
We obtain the spontaneous polarization of the LCO phase by comparing the polarization change between two symmetry variants related by the operation $(\bar{x},y+1/2,\bar{z})$. Assuming the change arises primarily from uniform transfer of electrons from V$^{3+}$ to V$^{4+}$, an estimate with a linearized expression for the polarization using nominal valences for the ions gives 29 $\mu$C/cm$^{2}$ for the transfer across the LaO plane and 22  $\mu$C/cm$^{2}$ across the SrO plane. The corresponding values obtained by Berry phase computed  with an appropriate branch choice are 34 and 17 $\mu$C/cm$^{2}$, respectively. 

\begin{figure}[htbp]
\begin{center}
\includegraphics[width=0.9\columnwidth, angle=-0]{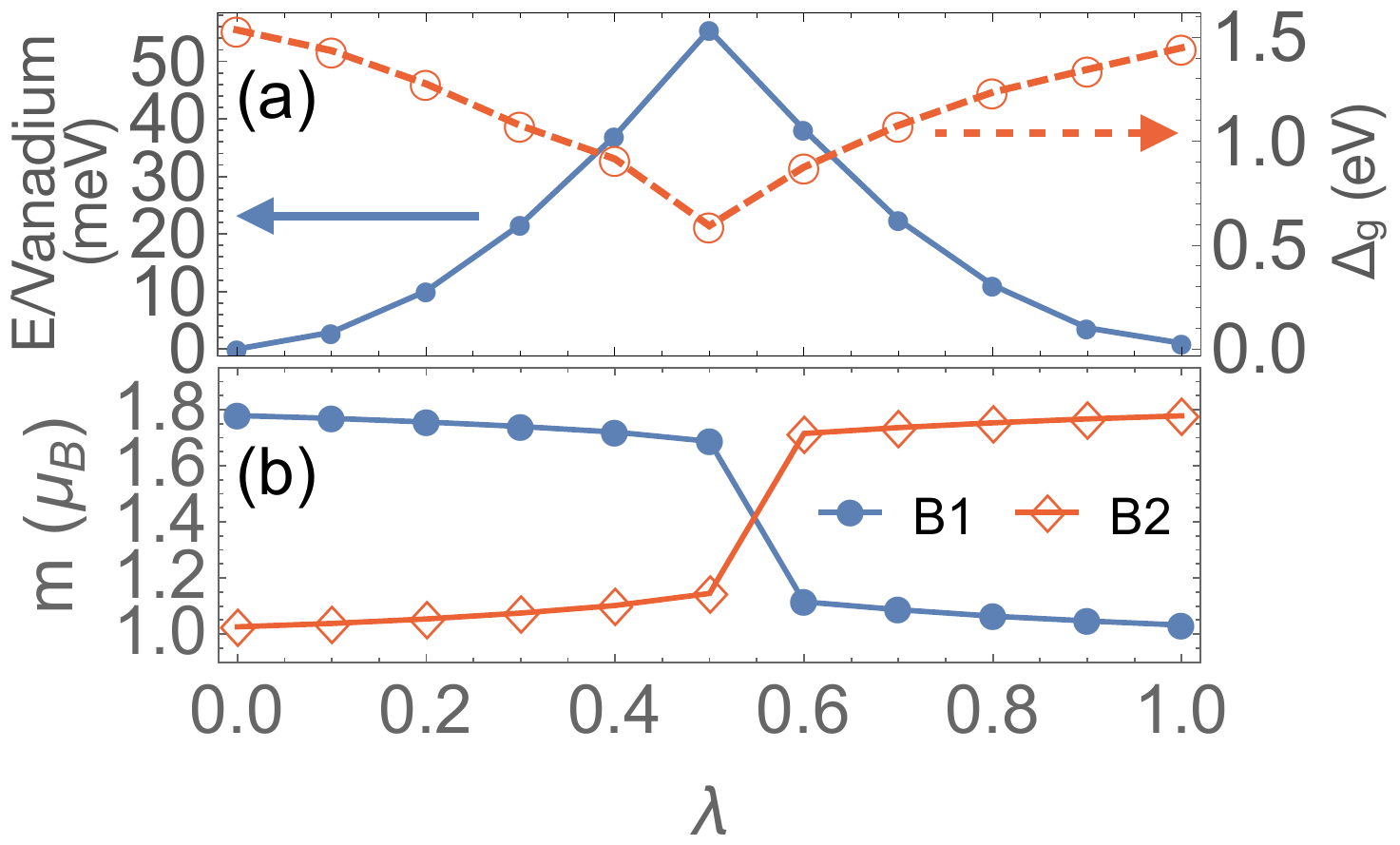}
\caption{(Color online) (a) Total energy per vanadium (blue solid line) and band gap (red dashed line) as a function of the interpolation parameter $\lambda$. The atomic arrangement of SCO phase ($\lambda=0$) is linearly interpolated to that of LCO phase ($\lambda=1$). (b) Magnetic moments of V cations vs $\lambda$. The symbols represent the position of V-cations defined in Fig.~\ref{fig:copatterns}.}
\label{fig:strintp}
\end{center}
\end{figure}

The epitaxial strain dependence of the energies of the SCO and LCO phases with C-AFM magnetic ordering is shown in Fig.~\ref{fig:envsstrain}. We find that the energy difference increases with compressive strain, while for tensile strain above 1\%, it is close to zero. This suggests that in films under tensile strain, the polar LCO phase could be induced by an applied electric field, and a antiferroelectric double hysteresis loop could be observed. It is also possible that if the polar phase does not switch back to the nonpolar phase on removal of the field, the system would exhibit a pseudo-ferroelectric hysteresis loop. 

To investigate the energetics of the coupling between charge ordering and crystal structure, we computed the energy, band gap, and magnetic moments for structures obtained from a linear interpolation between the SCO ($\lambda=0$) and the LCO ($\lambda=1$) phase at 1\% tensile strain, with results shown in Fig.~\ref{fig:strintp}. Along the path starting from each endpoint phase, the energy increases smoothly and the band gap decreases from about 1.5 eV down to a minimum of 0.55 eV. At about the midway point, there is a pronounced cusp in the energy, with a calculated energy barrier along this path of about 55 meV per vanadium.  On each side of the cusp, there is little change in the magnetic moments of the V cations at B1 and B2 positions (Fig.~\ref{fig:copatterns}), which we take as indicators of their charge state. As the cusp is crossed, the B1 and B2 magnetic moments exchange. This suggests that the charge disproportionation drives the structural relaxation, rather than vice versa.

The first-principles results presented here shown that in the (LaVO$_3$)$_{1}$/(SrVO$_3$)$_{1}$ superlattice, a layered charge-ordering pattern combines with the symmetry breaking by cation layering to produce a low-energy polar phase with a substantial polarization, comparable to that of the prototypical ferroelectric BaTiO$_3$. While this polar phase is not the predicted ground state, its energy is low enough so that it can be stabilized by an applied electric field. We note that the computed energy differences between the various charge order patterns considered are much smaller than the classical electrostatic energy differences between corresponding arrangements of 3+ and 4+ point charges in a fixed compensating background. This suggests that there is substantial screening by the other charges in the system, for example by charge rearrangement in the apical oxygens, which are shared by 3+ and 4+ centered octahedra in all the arrangements considered here. These low energy differences also suggest that while the tendency for the vanadiums to disproportionate into V$^{3+}$ and V$^{4+}$ is very strong, the critical temperature for long range charge order in the ground-state nonpolar phase should be very low, while the system can be easily ordered into the polar phase by application of an electric field. The energy relative to the nonpolar phases, and therefore the critical electric field, can be lowered in thin films under tensile strain.

The large spontaneous polarization of the layered phase arises from electron transfer from V$^{3+}$ to V$^{4+}$  ions that transforms one symmetry variant of the polar phase into the other. A branch choice corresponding to  minimal rearrangement of charges and ions has been established as a good predictor of the measured switching polarization for known ferroelectrics \cite{John16}. The present case is unusual in that two equally minimal rearrangements (electron transfer across the LaO plane and across the SrO plane) give values for the spontaneous polarizations that differ by more than a factor of two. These and other subtleties of the polarization will be further discussed elsewhere \cite{Park16}.

The stabilization of a polar phase by the combination of cation layering and layered charge ordering is expected to operate in other transition metal oxide superlattices, the key ingredient being stable multiple valence states with an average valence controlled by the A cations and a mechanism for screening the electrostatic energy differences of the relevant point charge arrangements. First-principles high-throughput searches should be useful in identifying promising candidate systems for further theoretical and experimental investigation.

In summary, we predict a charge-order-driven polar phase in the 1:1 superlattice composed of perovskite oxides, LaVO$_{3}$ and SrVO$_{3}$. We find three low-energy antiferromagnetic Mott-insulating states with distinct charge ordered patterns; the SCO and CCO phases with zero out-of-plane polarization and the LCO phase with a sizable out-of-plane polarization. We find that 
the out-of-plane spontaneous polarization of the polar phase is comparable to that of conventional ferroelectric perovskite oxides and is much larger than in other charge-order-driven ferroelectrics. It can be understood as a transfer of electrons between V ions in the transformation between up and down polarization states.  The energy difference between the polar phase and ground-state nonpolar phase is small and can be controlled by (001) epitaxial strain. This suggests that the polar phase can be induced by an applied electric field and that the critical field will decrease under tensile strain. Such phases are expected also to occur in other layered superlattices with transition metal ions stable in multiple valence states.

\begin{acknowledgments}
We thank C. Dreyer, R. Engel-Herbert, V. Gopalan, D. R. Hamann, H. N. Lee, J. Liu, B. Monserrat, D. Vanderbilt, M. Ye, and Y. Zhou, for valuable discussion. First-principles calculations were performed on the Rutgers University Parallel Computer (RUPC) cluster. This work is supported by ONR N00014-11-1-0666 and ONR N00014-12-0261.
\end{acknowledgments}

\bibliography{Park_refs.bib}
\end{document}